\documentclass[12pt]{iopart}
\usepackage{graphics}
\usepackage{graphicx}
\usepackage[sort&compress,numbers]{natbib}

\begin{document}

\title[Electronic and Structural Properties of MoSSe/MoX$_2$ (X=S,Se) Heterojunctions]{Electronic and Structural Properties of Janus SMoSe/MoX$_2$ (X=S,Se) In-plane Heterojunctions: A DFT Study}

\author{Ramiro M. dos Santos$^1$, Wiliam F. da Cunha$^1$, William F. Giozza$^2$, Rafael T. de Sousa J\'unior$^2$, Luiz F. Roncaratti J\'unior$^1$, and Luiz A. Ribeiro J\'unior$^{1,*}$}

\address{$^{1}$ \quad Institute of Physics, University of Bras\'ilia, Bras\'ilia, 70910-900, Brazil}
\address{$^{2}$ \quad Department of Electrical Engineering, University of Bras\'{i}lia 70919-970, Brazil}
\ead{ribeirojr@unb.br}
\vspace{10pt}
\begin{indented}
\item[]December 2020
\end{indented}

\begin{abstract}
The electronic and structural properties of Janus MoSSe/MoX$_2$ (X=S,Se) in-plane heterojunctions, endowed with single-atom vacancies, were studied using density functional theory calculations. The stability of these structures was verified from cohesion energy calculations. Results showed that single-atom vacancies induce the appearance of flat midgap states, and a substantial amount of charge is localized in the vicinity of these defects. As a consequence, these heterojunctions presented an intrinsic dipole moment. No bond reconstructions were noted by removing an atom from the lattice, regardless of its chemical species. Our calculations predicted indirect electronic bandgap values between 1.6-1.7 eV.  
\end{abstract}

\section{Introduction}

Two-dimensional nanomaterials are the novel promising candidates in developing more efficient optoelectronic applications \cite{tan2017recent,koski2013new}. The advent of graphene was responsible for the rise of interest in these materials \cite{novoselov2004electric,geim2007rise,geim2009graphene}. Its unique mechanical and electronic properties have impacted the field of nanoelectronics \cite{chang2013graphene}. However, the null bandgap of graphene bandgap is a drawback in this field. Since a semiconducting bandgap is crucial to the design of more efficient optoelectronic devices, the searching for systems with similar properties to graphene but that might present a bandgap is fundamental.

Recently, MX$_2$ transition metal dichalcogenides (TMDs, where X is a chalcogen and M a transition metal atom) \cite{splendiani2010emerging,li2012bulk}, hexagonal boron \cite{song2010large}, and aluminum nitride \cite{taniyasu2006aluminium}, among other possibilities of 2D systems \cite{mohammad1995emerging,manzeli20172d,anasori20172d}, have been investigated. The use of these materials in renewable energy engineering has resulted in advances in photovoltaic and energy storage applications \cite{das2019role,zhang20162d}. Interfaces built with these materials are alternatives that can be used in manufacturing these devices \cite{novoselov20162d,geim2013van}. The large-scale synthesis of these materials may lead to the appearance of local defects, such as substitutional doping and vacancies \cite{lin20162d}. These defects impact the electronic properties of the materials and may result in desirable consequences for their application in nanoelectronics, as the bandgap tuning \cite{feng2017doping}, for instance. 

Janus MoSSe monolayer (from now on MoSSe) is structurally similar to MoX$_2$ (X=S,Se), but the former contains two distinct planes of chalcogen atoms, one composed of sulfur and another one of selenium atoms \cite{lu2017janus,zhang2017janus} (see Figure \ref{fig:systems}). The lattice parameters between MoSSe and MoX$_2$ differ by 0.02 \r{A} \cite{articleJMC}, which allows their vertical packing \cite{zhang2020enhancement} or in-plane assembly \cite{guo2020designing}. Several experimental and theoretical works investigated the electronic and structural properties of Van der Waals heterostructures of MoSSe$/$MoS$_2$ monolayers \cite{zhang2020enhancement,bafekry2020van,riis2018efficient,li2019intrinsic,wang2020optical,zhao2020van}. Nevertheless, studies about the MoSSe/MoS$_2$ in-plane heterojunction are very few in the literature \cite{guo2020designing}, and its electronic structure in the presence of lattice defects was not studied so far.   

Herein, the electronic and structural properties of MoSSe/MoX$_2$ in-plane heterojunctions were studied using density functional theory (DFT) calculations. The single-atom vacancies were considered by removing one atom of each atomic species. Our model systems are based in two distinct in-plane interfaces with single-atom vacancies: MoSSe$/$MoS$_2$ and MoSSe$/$MoSe$_2$. For comparison purposes, the results obtained for these systems are contrasted with the ones for their related non-defective heterojunctions. Results suggested that the defective heterojunctions are stable and may present an intrinsic dipole moment. 

\section{Details of Modeling}

\begin{figure}[!htb]
    \centering
    \includegraphics[width=0.8\linewidth]{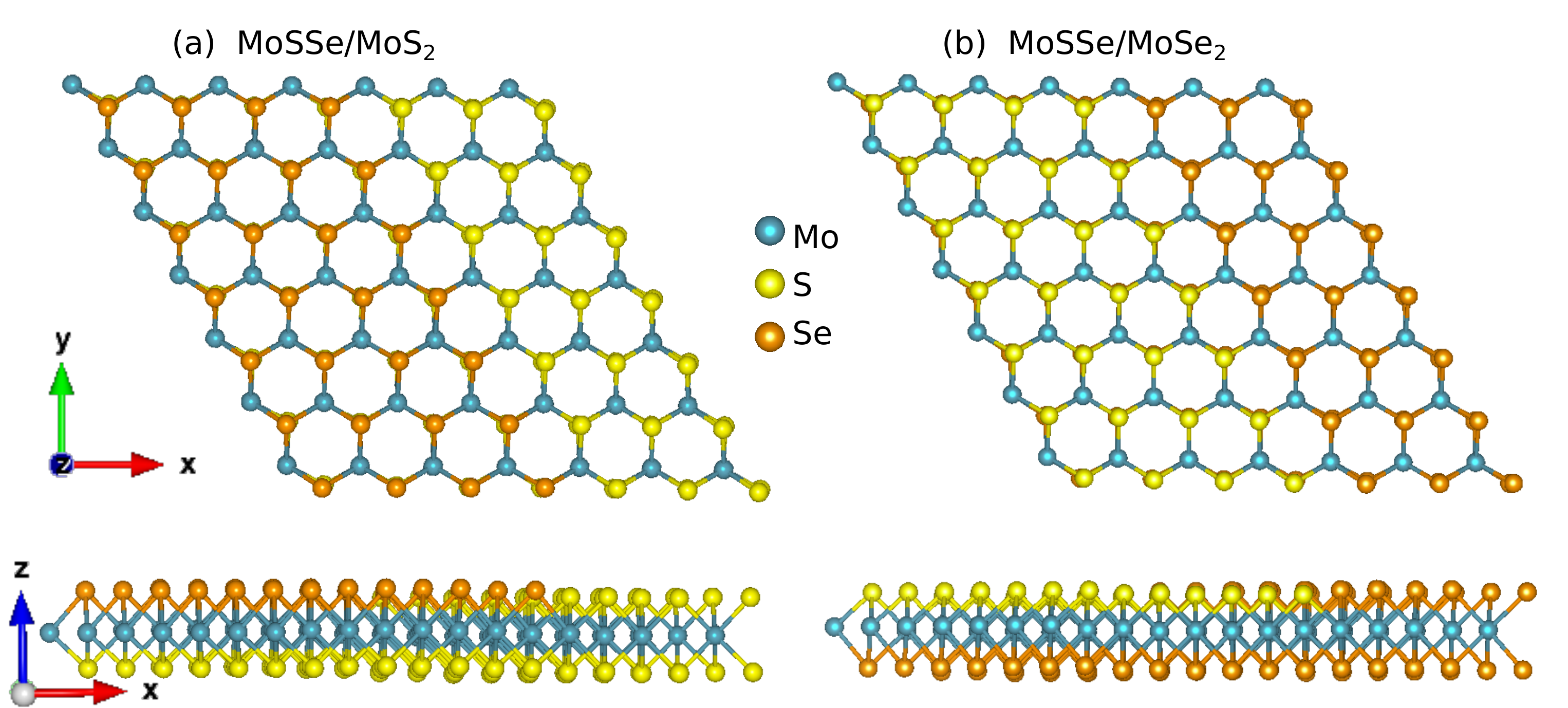}
    \caption{Schematic representation of the model pristine (a) MoSSe/MoS$_2$ and (b) MoSSe/MoSe$_2$ structures studied here. The top and bottom panels illustrate the top and side views of the lattice, respectively.}
    \label{fig:systems}
\end{figure}

The electronic and structural properties of Mosse/MoX$_2$ in-plane heterojunctions were studied using DFT calculations as implemented in the SIESTA code \cite{soler2002siesta}. These calculations were performed within the scope of generalized gradient approximation (GGA) with localized basis sets \cite{PhysRevLett.77.3865}. The exchange-correlation functional used is based on the Perdew–Burke–Ernzerhof (PBE) framework \cite{PhysRevLett.80.891}. To treat the electron core interaction, we used the Troullier–Martins norm-conserving pseudopotentials \cite{PhysRevB.64.235111}. We also include polarization effects and the Kohn–Sham orbitals are expanded with double-$\zeta$ basis \cite{PhysRevB.64.235111}.  The structural relaxation of all model lattices studied here is carried out until the force on each atom is less than $10^{-3}$ eV/\r{A}, and the total energy change between each self-consistency step achieved a value less or equal to $10^{-5}$ eV. The Brillouin zone is sampled by a fine $9\times 9\times 3$ grid and to determine the self-consistent charge density we use a mesh cutoff of 400 Ry, and the ground state structure was obtained after the total forces on each atom reached the value of 0.001 eV/\r{A}. A supercell geometry was adopted with a vacuum distance of 30 \r{A} to avoid interaction among each structure and its images. Figure \ref{fig:systems} illustrates the model pristine lattices studied here. Importantly, the vacancy endowed ones are presented below. Here, we considered interfaces where the size of the MoSSe lattice is predominant. The idea is to investigate how the electronic properties are affected by a small change in the concentration ratio between sulfur and selenium atoms.

\section{Results}

\begin{figure}[!htb]
    \centering
    \includegraphics[width=0.8\linewidth]{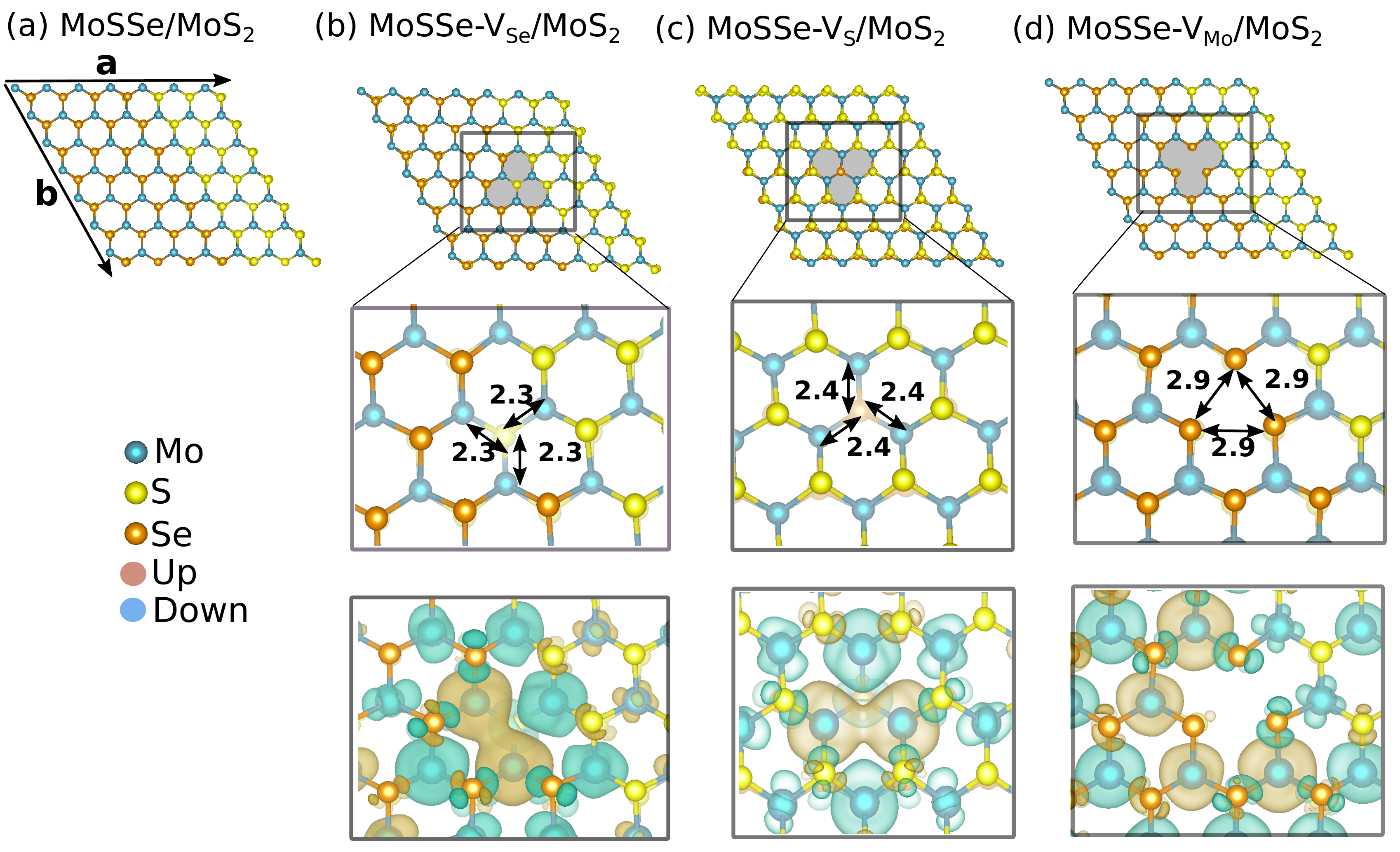}
    \caption{Schematic representation of the model supercells (top panels), the bond configuration (middle panels), and the charge localization (bottom panels) in the vicinity of the defects. (a) shows the model lattice for the pristine case. Panels (b), (c), and (d) depict, respectively, the bond configuration and charge density localization for the cases with a selenium vacancy (MoSSe-V$_{Se}$/MoS$_2$), a sulfur vacancy (MoSSe-V$_S$/MoS$_2$), and a molybdenum vacancy (MoSSe-V${_{Mo}}$/MoS$_2$).}
    \label{fig:bondcharge01}
\end{figure}

We begging our discussions by presenting the bond configuration and charge density localization of the MoSSe/MoS$_2$ lattices containing single-atom vacancies. In Figure \ref{fig:bondcharge01}, the top, middle, and bottom sequence of panels illustrate the model supercells, the bond configuration, and the charge localization in the vicinity of the defects, respectively. Figure \ref{fig:bondcharge01}(a) shows the model lattice for the pristine case (MoSSe/MoS$_2$). Figures \ref{fig:bondcharge01}(b), \ref{fig:bondcharge01}(c), and \ref{fig:bondcharge01}(d) depict, respectively, the bond configuration and charge density localization for the cases with a selenium vacancy (MoSSe-V$_{Se}$/MoS$_2$), a sulfur vacancy (MoSSe-V$_S$/MoS$_2$), and a molybdenum vacancy (MoSSe-V${_{Mo}}$/MoS$_2$). The atoms were removed from regions nearby the interface. In the middle panels, one can observe that the bond lengths between the atoms in the vacancy region formed by removing a Se (Figure \ref{fig:bondcharge01}(b)) or S (Figure  \ref{fig:bondcharge01}(c)) atoms are 2.3 and 2.4, respectively \r{A}. These values slightly deviate from what was reported for a Mo-S bond at homogeneous MoS$_2$ monolayers, which is about 2.4 \r{A} \cite{articleJMC}. In the case of Mo vacancy (Figure  \ref{fig:bondcharge01}(d)), the Se-Se distances within the vacancy are 2.9 \r{A}. In the rest of the lattice, the related distances are 3.1 \r{A}. It is worthwhile to stress that no bond reconstructions take place, and the defective region retained its symmetry even after ground-state calculations. 

The bottom sequence of panels in Figure \ref{fig:bondcharge01} presents the charge density for spin-up (red spots) and spin-down (blue spots) electrons, in the vicinity of the vacancy, for the related defective lattices. One can realize that in the MoSSe-V$_{Se}$/MoS$_2$ and MoSSe-V$_{S}$/MoS$_2$ cases, there is an equivalent distribution of both spin channels over the Mo atoms nearby the defect. On the other hand, in the MoSSe-V$_{Mo}$/MoS$_2$ case, we observed the prevalence of non-binding states over Mo atoms surrounding the vacancy. In this case, there is a substantial distribution of only spin-up electrons. Such a charge configuration contributes to increasing the intrinsic dipole moment in this particular case (see Table \ref{tab1}).

\begin{figure}[!htb]
    \centering
    \includegraphics[width=0.8\linewidth]{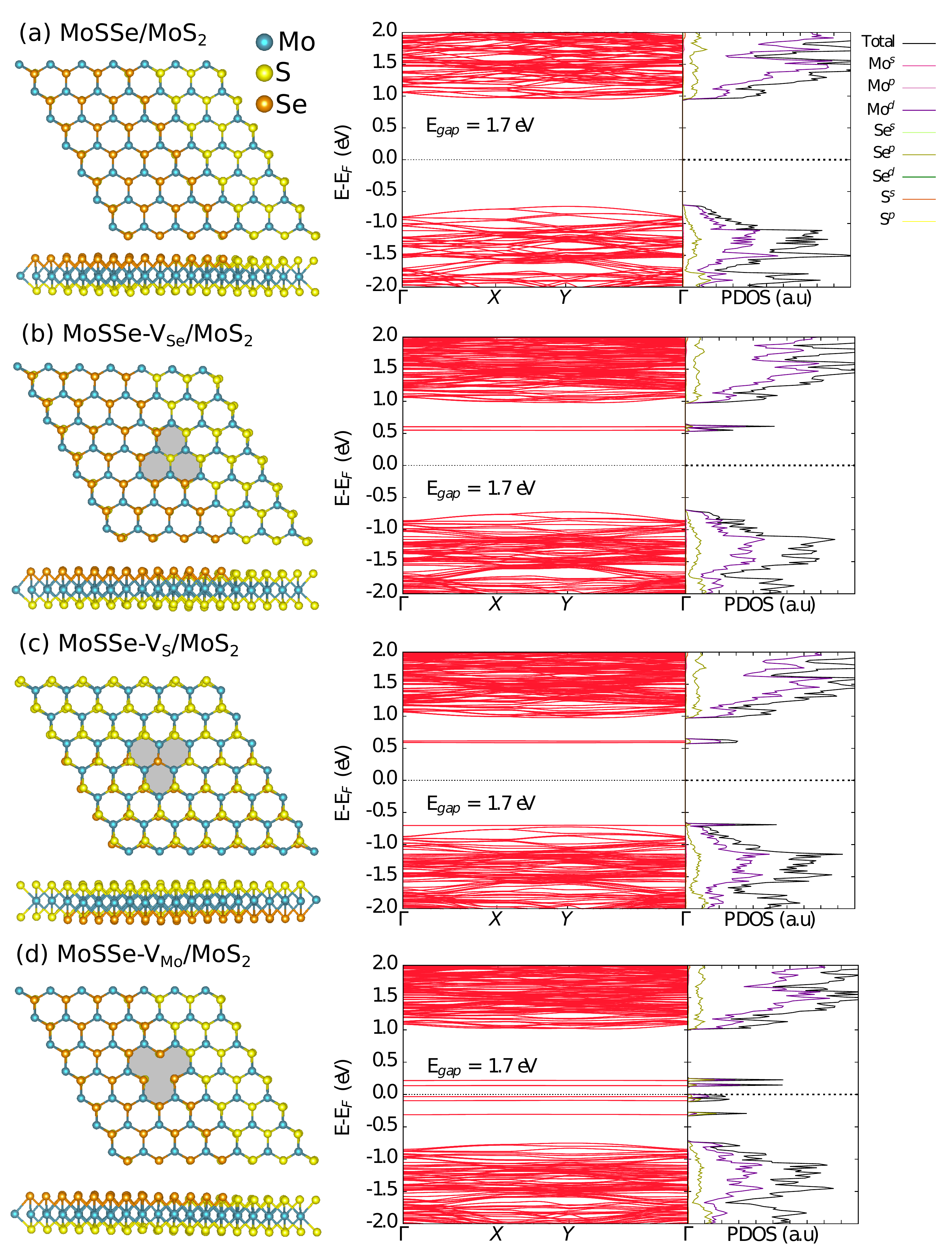}
    \caption{Band structure and related partial density of states (PDOS) for the cases presented in Figure \ref{fig:bondcharge01}. (a) shows these results for the pristine case (MoSSe/MoS$_2$). Panels (b), (c), and (d) depict, respectively, the bond configuration and charge density localization for the cases with a selenium vacancy (MoSSe-V$_{Se}$/MoS$_2$), a sulfur vacancy (MoSSe-V$_S$/MoS$_2$), and a molybdenum vacancy (MoSSe-V${_{Mo}}$/MoS$_2$).}
    \label{fig:banddos01}
\end{figure}

Figure \ref{fig:banddos01} shows the band structure and related partial density of states (PDOS) for the cases presented in Figure \ref{fig:bondcharge01}. All the systems have presented an indirect bandgap of 1.7 eV. This value is 0.3 eV lower than one for homogeneous MoS$_2$, which is approximately 2.0 eV \cite{splendiani2010emerging}. The bandgap value was not altered in the presence of single-atom vacancies. Instead, it was observed the arising of flat midgap sates by removing an atom, and the PDOS showed a major contribution of the $d$-orbital of molybdenum for these intragap states.

In the MoSSe-V$_{Se}$/MoS$_2$ case, there are three molybdenum atoms bonded to one sulfur atom within the vacancy region with a bond length of 2.3 \r{A} (see Figure \ref{fig:bondcharge01}(b)). This type of vacancy leads to the appearance of three dangling bonds related to the molybdenum atoms. The electrons belonging to these dangling bonds occupy the $d$-orbital of the molybdenum atom. Three of them are involved in covalent bonds with the electrons of the $p$-orbital of the sulfur atom. The remaining electrons in the dangling bonds are characterized by the flat midgap states. This trend is also observed in the MoSSe-V$_{S}$/MoS$_2$ case. In the MoSSe-V$_{Mo}$/MoS$_2$ heterojunction, the flat midgap states refer to the electrons in the dangling bonds of molybdenum and selenium atoms in the vicinity of the vacancy, as suggested by the PDOS in Figure \ref{fig:banddos01}(d). Note that the $p$-orbital of selenium and the $d$-orbital of molybdenum have major contributions to these states.      

\begin{figure}[!htb]
    \centering
    \includegraphics[width=0.8\linewidth]{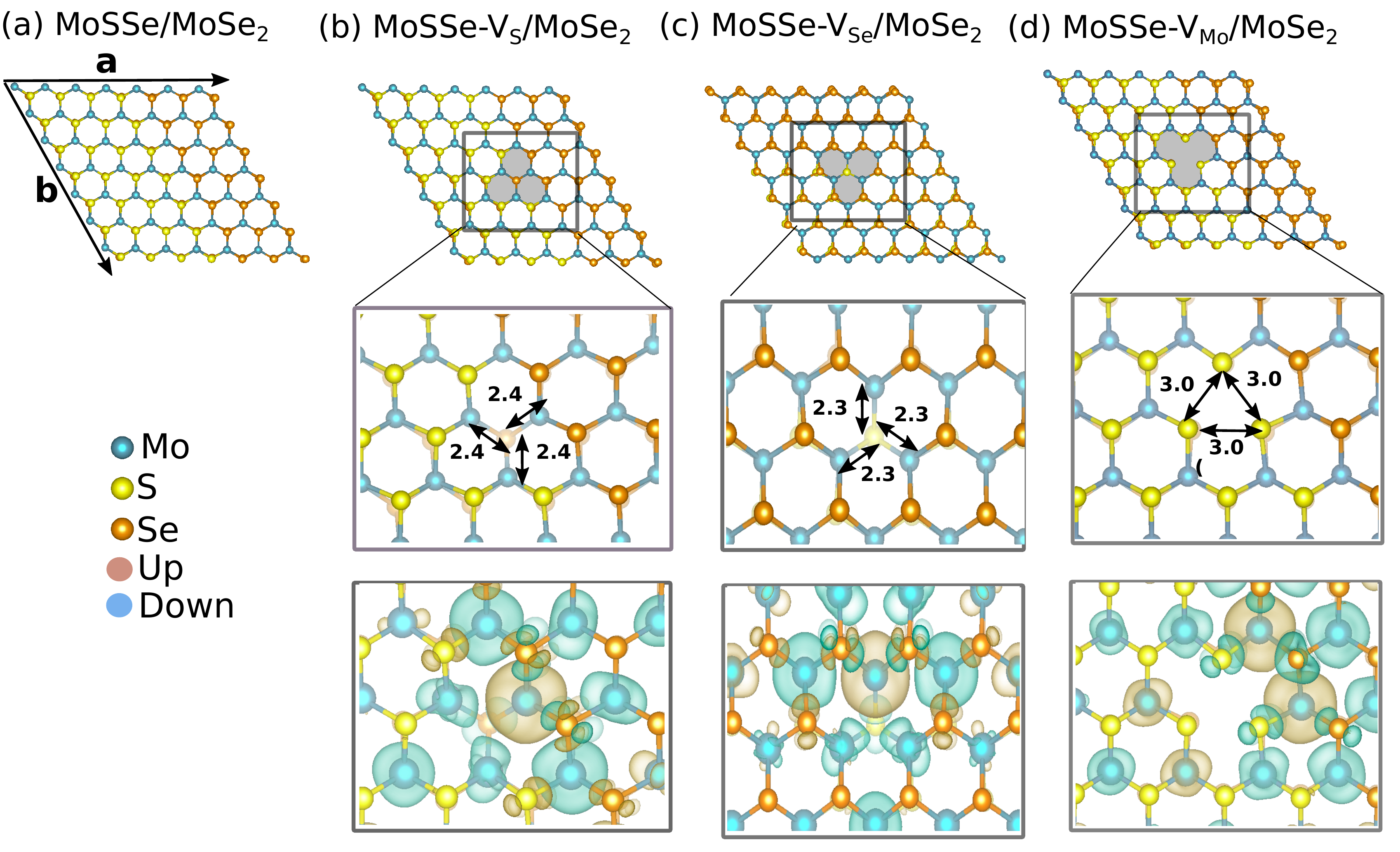}
    \caption{Schematic representation of the model supercells (top panels), the bond configuration (middle panels), and the charge localization (bottom panels) in the vicinity of the defects. (a) shows the model lattice for the pristine case (MoSSe/MoSe$_2$). Panels (b), (c), and (d) depict, respectively, the bond configuration and charge density localization for the cases with a selenium vacancy (MoSSe-V$_{Se}$/MoSe$_2$), a sulfur vacancy (MoSSe-V$_S$/MoSe$_2$), and a molybdenum vacancy (MoSSe-V${_{Mo}}$/MoSe$_2$).}
    \label{fig:bondcharge02}
\end{figure}

\begin{figure}[!htb]
    \centering
    \includegraphics[width=0.8\linewidth]{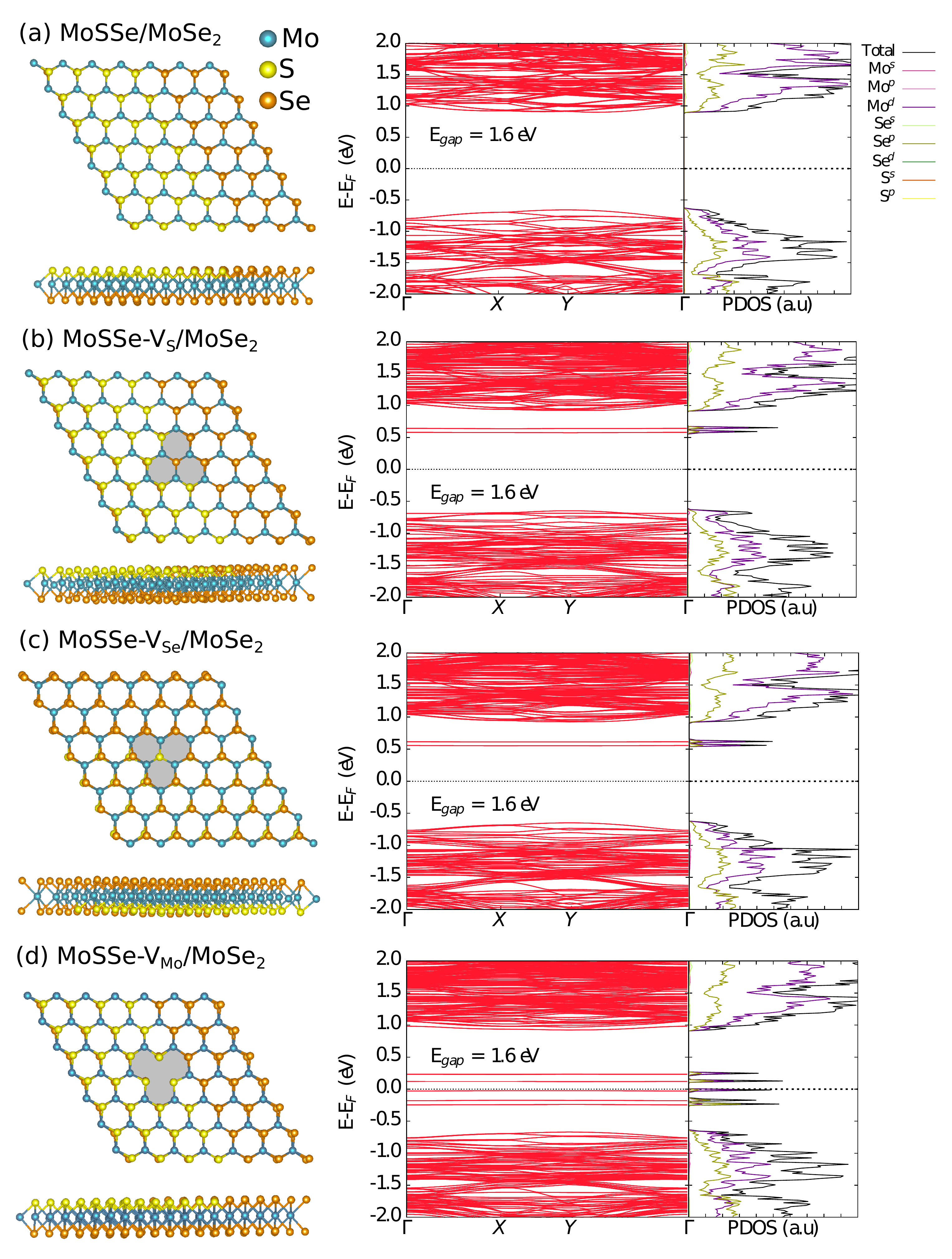}
    \caption{Band structure and related partial density of states (PDOS) for the cases presented in Figure \ref{fig:bondcharge02}. (a) shows these results for the pristine case (MoSSe/MoSe$_2$). Panels (b), (c), and (d) depict, respectively, the bond configuration and charge density localization for the cases with a selenium vacancy (MoSSe-V$_{Se}$/MoSe$_2$), a sulfur vacancy (MoSSe-V$_S$/MoSe$_2$), and a molybdenum vacancy (MoSSe-V${_{Mo}}$/MoSe$_2$).}
    \label{fig:banddos02}
\end{figure}

Now, we present similar results for MoSSe/MoSe$_2$ lattices containing single-atom vacancies. The resulting systems present a supercell with a larger concentration of Se atoms. Figure \ref{fig:bondcharge02} shows the geometry configurations and charge density localization for these systems.  Figure \ref{fig:bondcharge02}(a) shows the model lattice for the pristine case (MoSSe/MoSe$_2$). Figures \ref{fig:bondcharge02}(b), \ref{fig:bondcharge02}(c), and \ref{fig:bondcharge02}(d) depict, respectively, the bond configuration and charge density localization for the cases with a selenium vacancy (MoSSe-V$_{Se}$/MoSe$_2$), a sulfur vacancy (MoSSe-V$_S$/MoSe$_2$), and a molybdenum vacancy (MoSSe-V${_{Mo}}$/MoSe$_2$). Once again, the top panels show a schematic representation of the resulting system, the middle ones present the lattice bond lengths nearby the vacancy, and the bottom panels illustrate the charge density localization. The atoms were removed from regions nearby the interface. In the middle panels, one can observe that the bond lengths between the atoms in the vacancy region formed by removing a Se (Figure \ref{fig:bondcharge02}(b)) or S (Figure  \ref{fig:bondcharge02}(c)) atoms are 2.4 and 2.3, respectively \r{A}. As mentioned above, these values slightly deviate from what was reported for a Mo-S bond at homogeneous MoS$_2$ monolayers (2.4 \r{A} \cite{articleJMC}). By contrasting these results with the ones presented in Figure \ref{fig:bondcharge01}, one can note that the bond length configuration was not sensitive to a small change in the concentration ratio between sulfur and selenium atoms. In the case of Mo vacancy (Figure  \ref{fig:bondcharge02}(d)), the Se-Se distances within the vacancy are 3.0 \r{A}. In the rest of the lattice, the related distances are 3.1 \r{A}. These bond configuration is similar to what was shown in Figure \ref{fig:bondcharge02}(d). As expected, a small change in the concentration ratio between sulfur and selenium atoms has not impacted the bond configuration also in the MoSSe-V${_{Mo}}$/MoSe$_2$) case. No bond reconstructions were observed in all systems presented in Figure \ref{fig:bondcharge02} upon ground sate calculation.

Figure \ref{fig:banddos02} presents the band structure and related partial density of states (PDOS) for the cases presented in Figure \ref{fig:bondcharge02}. The band structure for the MoSSe$/$MoSe$_2$ heterojunctions are very similar to the MoSSe$/$MoS$_2$ ones (see Figure \ref{fig:banddos01}). In this sense, a small change in the concentration ratio between sulfur and selenium atoms has not impacted the electronic properties. All the MoSSe$/$MoSe$_2$ heterojunctions presented an indirect bandgap of 1.6 eV. The bandgap value was not altered in the presence of single-atom vacancies, and the arising of flat midgap sates were also observed. As in the vacancy-endowed MoSSe/MoS$_2$ cases, in the MoSSe$/$MoSe$_2$ ones, the PDOS showed a major contribution of the $d$-orbital of molybdenum for these midgap states.

Finally, Table \ref{tab1} presents the values obtained for the cohesion energy and electric dipole moment related to the model MoSSe$/$MoS$_2$ and MoSSe$/$MoSe$_2$ heterojunctions presented above. The energetic stability of these structures were measured using the cohesion energy formula, that is expressed as
\begin{equation}
\displaystyle
E_{coh}=\frac{E_{total}-(E_{Mo} \cdot X_{Mo}+E_S \cdot X_S+E_{Se} \cdot X_{Se})}{X_{total}},
\end{equation}
\noindent where $E_{total}$ and $X_{total}$ are the total energy and the total number of atoms, and $X_v$ and $E_v$ are the number of atoms and total energy of an isolated atom $v$, respectively. In this case, $v$ stands for $Mo$, $S$, or $Se$. It can be observed in Table \ref{tab1} that the inclusion of a single-atom vacancy or a small change in the concentration ratio between sulfur and selenium atoms does not impact the cohesion energy and dipole moment values. One can note that the cohesion energy values for the defective lattices are very close to the ones for the pristine cases. These results suggest that model defective lattices are energetically stable. Moreover, all the systems presented an intrinsic dipole moment. 

\begin{table}[!htb]
\scriptsize
\centering
   \begin{tabular}{|c|c|c|c|c|}
\hline
               & MoSSe/MoS$_2$ & MoSSe-V$_{Se}$/MoS$_2$ & MoSSe-V$_{S}$/MoS$_2$ & MoSSe-V$_{Mo}$/MoS$_2$\\
\hline               
  E$_{coh}$ (eV)     & -6.8 & -6.8 & -6.8 & -6.7   \\ 
  \textbf{P} (Debye)    & -3.8 & -4.0  &  -3.8 & -3.8   \\  
    \textbf{a} (\AA) &21.60 & 21.91 & 21.91 & 21.96 \\
    \textbf{b} (\AA) &21.60 & 21.91 & 21.94 & 22.09 \\
\hline
               & MoSSe$_2$/MoSe$_2$ & MoSSe$_2$-V$_{Se}$/MoSe$_2$ & MoSSe-V$_{S}$/MoSe$_2$ & MoSSe-V$_{Mo}$/MoSe$_2$\\

\hline
  E$_{coh}$ (eV)     & -6.8 & -6.8 & -6.8 & -6.7   \\
    \textbf{P}  (Debye)    & -3.8 & -3.7  &  -3.8 & -3.7   \\  
 \textbf{a}  (\AA) &21.96 & 21.91 & 21.91 & 21.96 \\
     \textbf{b} (\AA) &21.98 & 21.94 & 21.92 & 22.02 \\
\hline
\end{tabular}
\caption{Cohesion energy ($E_{coh}$), electric dipole moment ($\mathbf{P}$) and, lattice parameters $\mathbf{a}$ and $\mathbf{b}$.}
\label{tab1}
 \end{table}

\section{Conclusion} 

In summary, the electronic and structural properties of MoSSe/MoX$_2$ (X=S,Se) in-plane heterojunctions in the presence of single-atom (sulfur, selenium, or molybdenum) vacancies were investigated in the framework of density functional theory calculations. Results showed that no bond reconstructions were noted by removing an atom from the lattice. Our calculations predicted indirect electronic bandgap values between 1.6-1.7 eV, and the presence of an intrinsic dipole moment in the defective lattices. As a general trend, we concluded that the inclusion of a single-atom vacancy or a small change in the concentration ratio between sulfur and selenium atoms does not impact the electronic and structural properties of the model MoSSe/MoX$_2$ heterojunctions studied here.

\section{Acknowledgments}

The authors gratefully acknowledge the financial support from Brazilian Research Councils CNPq, CAPES, and FAPDF and CENAPAD-SP for providing the computational facilities. W.F.G. gratefully acknowledges the financial support from FAP-DF grant $0193.0000248/2019-32$. L.A.R.J. gratefully acknowledges the financial support from CNPq grant 302236/2018-0. R.T.S.J. gratefully acknowledges, respectively, the financial support from CNPq grant $465741/2014-2$, CAPES grants $88887.144009/2017-00$, and FAP-DF grants $0193.001366/2016$ and $0193.001365/2016$. L.A.R.J. gratefully acknowledges the financial support from DPI/DIRPE/UnB (Edital DPI/DPG $03/2020$) grant $23106.057541/2020-89$ and from IFD/UnB (Edital $01/2020$) grant $23106.090790/2020-86$.  

\bibliographystyle{iopart-num}
\bibliography{references}

\providecommand{\newblock}{}
\begin{thebibliography}{10}
\expandafter\ifx\csname url\endcsname\relax
  \def\url#1{{\tt #1}}\fi
\expandafter\ifx\csname urlprefix\endcsname\relax\def\urlprefix{URL }\fi
\providecommand{\eprint}[2][]{\url{#2}}

\bibitem{tan2017recent}
Tan C, Cao X, Wu X~J, He Q, Yang J, Zhang X, Chen J, Zhao W, Han S, Nam G~H
  {\em et~al.\/} 2017 {\em Chemical reviews\/} {\bf 117} 6225--6331

\bibitem{koski2013new}
Koski K~J and Cui Y 2013 {\em Acs Nano\/} {\bf 7} 3739--3743

\bibitem{novoselov2004electric}
Novoselov K~S, Geim A~K, Morozov S~V, Jiang D, Zhang Y, Dubonos S~V, Grigorieva
  I~V and Firsov A~A 2004 {\em science\/} {\bf 306} 666--669

\bibitem{geim2007rise}
Geim A~K and Novoselov K~S 2007 {\em Nature Materials\/} {\bf 6} 183--191

\bibitem{geim2009graphene}
Geim A~K 2009 {\em science\/} {\bf 324} 1530--1534

\bibitem{chang2013graphene}
Chang H and Wu H 2013 {\em Advanced Functional Materials\/} {\bf 23} 1984--1997

\bibitem{splendiani2010emerging}
Splendiani A, Sun L, Zhang Y, Li T, Kim J, Chim C~Y, Galli G and Wang F 2010
  {\em Nano letters\/} {\bf 10} 1271--1275

\bibitem{li2012bulk}
Li H, Zhang Q, Yap C~C~R, Tay B~K, Edwin T~H~T, Olivier A and Baillargeat D
  2012 {\em Advanced Functional Materials\/} {\bf 22} 1385--1390

\bibitem{song2010large}
Song L, Ci L, Lu H, Sorokin P~B, Jin C, Ni J, Kvashnin A~G, Kvashnin D~G, Lou
  J, Yakobson B~I {\em et~al.\/} 2010 {\em Nano letters\/} {\bf 10} 3209--3215

\bibitem{taniyasu2006aluminium}
Taniyasu Y, Kasu M and Makimoto T 2006 {\em Nature\/} {\bf 441} 325--328

\bibitem{mohammad1995emerging}
Mohammad S~N, Salvador A~A and Morkoc H 1995 {\em Proceedings of the IEEE\/}
  {\bf 83} 1306--1355

\bibitem{manzeli20172d}
Manzeli S, Ovchinnikov D, Pasquier D, Yazyev O~V and Kis A 2017 {\em Nature
  Reviews Materials\/} {\bf 2} 17033

\bibitem{anasori20172d}
Anasori B, Lukatskaya M~R and Gogotsi Y 2017 {\em Nature Reviews Materials\/}
  {\bf 2} 1--17

\bibitem{das2019role}
Das S, Pandey D, Thomas J and Roy T 2019 {\em Advanced Materials\/} {\bf 31}
  1802722

\bibitem{zhang20162d}
Zhang X, Hou L, Ciesielski A and Samor{\`\i} P 2016 {\em Advanced Energy
  Materials\/} {\bf 6} 1600671

\bibitem{novoselov20162d}
Novoselov K, Mishchenko o~A, Carvalho o~A and Neto A~C 2016 {\em Science\/}
  {\bf 353}

\bibitem{geim2013van}
Geim A~K and Grigorieva I~V 2013 {\em Nature\/} {\bf 499} 419--425

\bibitem{lin20162d}
Lin Z, McCreary A, Briggs N, Subramanian S, Zhang K, Sun Y, Li X, Borys N~J,
  Yuan H, Fullerton-Shirey S~K {\em et~al.\/} 2016 {\em 2D Materials\/} {\bf 3}
  042001

\bibitem{feng2017doping}
Feng S, Lin Z, Gan X, Lv R and Terrones M 2017 {\em Nanoscale Horizons\/} {\bf
  2} 72--80

\bibitem{lu2017janus}
Lu A~Y, Zhu H, Xiao J, Chuu C~P, Han Y, Chiu M~H, Cheng C~C, Yang C~W, Wei K~H,
  Yang Y {\em et~al.\/} 2017 {\em Nature nanotechnology\/} {\bf 12} 744--749

\bibitem{zhang2017janus}
Zhang J, Jia S, Kholmanov I, Dong L, Er D, Chen W, Guo H, Jin Z, Shenoy V~B,
  Shi L {\em et~al.\/} 2017 {\em ACS nano\/} {\bf 11} 8192--8198

\bibitem{articleJMC}
Yin W~J, Wen B, Wei X~L, Nie G and Liu L~M 2018 {\em Journal of Materials
  Chemistry C\/} {\bf 6}

\bibitem{zhang2020enhancement}
Zhang K, Guo Y, Ji Q, Lu A~Y, Su C, Wang H, Puretzky A~A, Geohegan D~B, Qian X,
  Fang S {\em et~al.\/} 2020 {\em Journal of the American Chemical Society\/}
  {\bf 142} 17499--17507

\bibitem{guo2020designing}
Guo Y, Lin Y, Xie K, Yuan B, Zhu J, Shen P~C, Lu A~Y, Su C, Shi E, Zhang K {\em
  et~al.\/} 2020 {\em arXiv preprint arXiv:2011.07690\/}

\bibitem{bafekry2020van}
Bafekry A, Yagmurcukardes M, Akgenc B, Ghergherehchi M and Nguyen C~V 2020 {\em
  Journal of Physics D: Applied Physics\/} {\bf 53} 355106

\bibitem{riis2018efficient}
Riis-Jensen A~C, Pandey M and Thygesen K~S 2018 {\em The Journal of Physical
  Chemistry C\/} {\bf 122} 24520--24526

\bibitem{li2019intrinsic}
Li F, Wei W, Wang H, Huang B, Dai Y and Jacob T 2019 {\em The journal of
  physical chemistry letters\/} {\bf 10} 559--565

\bibitem{wang2020optical}
Wang T, Su M, Jin H, Li J, Wan L and Wei Y 2020 {\em The Journal of Physical
  Chemistry C\/} {\bf 124} 15988--15994

\bibitem{zhao2020van}
Zhao H, Xie F, Liu Y, Bian B, Yang G, Ding Y, Gu Y, Yu Y, Zhang X, Huo X {\em
  et~al.\/} 2020 {\em Materials Science in Semiconductor Processing\/}  105588

\bibitem{soler2002siesta}
Soler J~M, Artacho E, Gale J~D, Garc{\'\i}a A, Junquera J, Ordej{\'o}n P and
  S{\'a}nchez-Portal D 2002 {\em Journal of Physics: Condensed Matter\/} {\bf
  14} 2745

\bibitem{PhysRevLett.77.3865}
Perdew J~P, Burke K and Ernzerhof M 1996 {\em Phys. Rev. Lett.\/} {\bf 77}(18)
  3865--3868
  \urlprefix\url{https://link.aps.org/doi/10.1103/PhysRevLett.77.3865}

\bibitem{PhysRevLett.80.891}
Perdew J~P, Burke K and Ernzerhof M 1998 {\em Phys. Rev. Lett.\/} {\bf 80}(4)
  891--891 \urlprefix\url{https://link.aps.org/doi/10.1103/PhysRevLett.80.891}

\bibitem{PhysRevB.64.235111}
Junquera J, Paz O, S\'anchez-Portal D and Artacho E 2001 {\em Phys. Rev. B\/}
  {\bf 64}(23) 235111
  \urlprefix\url{https://link.aps.org/doi/10.1103/PhysRevB.64.235111}

\end{thebibliography}

\end{document}